\documentclass[12pt]{article}
\pdfoutput=1
\usepackage{url} 
\usepackage{authblk}
\usepackage{amsmath}
\usepackage{subfig}
\usepackage{graphicx}
\usepackage{booktabs}
\usepackage{pdflscape} 
\usepackage{rotating} 
\usepackage{multirow}

\author[1]{J.~W.~Peterson\thanks{jw.peterson@inl.gov}}
\affil[1]{Idaho National Laboratory}

\newcommand{\bv}[1]{{\boldsymbol{#1}}}

\newcommand{\mean}[1]{\mu_{#1}}

\newcommand{\std}[1]{\sigma_{#1}}

\newcommand{\scaled}[1]{#1^{\ast}}

\title{Accurate curve fits of IAPWS data for high-pressure, high-temperature
single-phase liquid water based on the stiffened gas equation of
state}
\date{\today}

\begin{document}
\maketitle

\begin{abstract}
  We present a series of optimal (in the sense of least-squares) curve fits
  for the stiffened gas equation of state for single-phase liquid water.  At high
  pressures and (subcritical) temperatures, the parameters produced by these
  curve fits are found to have very small relative errors: less than 1\% in
  the pressure model, and less than 2\% in the temperature model.  At low
  pressures and temperatures, especially near the liquid-vapor transition
  line, the error in the curve fits increases rapidly.  The smallest
  pressure value for which curve fits are reported in the present work
  is 25 MPa, high enough to ensure that the fluid remains a single-phase
  liquid up to the maximum subcritical temperature of approximately 647K.
\end{abstract}

\section{Introduction\label{sec:introduction}}
The stiffened gas equation of state is frequently used in the simulation of
compressible liquids, in particular water~\cite{Berry_2010}, but can also be used
for simulating compressible gases including water vapor and air.  It is attractive primarily
due to the fact that it is a generalization of the ideal gas equation of state, and
because of its simplicity: only four independent constants ($\gamma$, $q$, $p_{\infty}$, and $c_v$)
are required to define the pressure and temperature models, which are given by
\begin{align}
  \label{eqn:stiffened_gas_p}
  p &= (\gamma-1)\rho(e-q) - \gamma p_{\infty}
  \\
  \label{eqn:stiffened_gas_T}
  T &= \frac{1}{c_v}\left(e-q - \frac{ p_{\infty} }{\rho}\right)
\end{align}
where $p$ is pressure, $T$ is temperature, $\rho$ is density, and $e$ is the internal energy
of the fluid.  We note that a fifth constant, denoted $q'$, is required to model the Gibbs free enthalpy,
however this quantity is not needed for performing typical thermohydraulic calculations.
The parameter $c_v$ used in~\eqref{eqn:stiffened_gas_T} should not be confused with
its typical meaning, i.e.\ the fluid's specific heat at constant volume.  Rather, we shall
treat it as a free parameter in the curve fitting process to be described subsequently.

For reference, the squared sound speed $c^2$, generalized adiabatic coefficient $\scaled{\gamma}$,
Gr\"{u}neisen coefficient $\Gamma$, and fundamental derivative $\mathcal{G}$~\cite{Menikoff_1989}
for the stiffened gas equation of state are given by
\begin{align}
  \label{eqn:c2}
  c^2 &= \frac{p + p_{\infty}}{\rho}
  \\
  \scaled{\gamma} &= \gamma \left( \frac{p + p_{\infty}}{p} \right)
  \\
  \Gamma &= \gamma-1
  \\
  \label{eqn:calG}
  \mathcal{G} &= \frac{1}{2}\left( \gamma+1  \right)
\end{align}
The quantities \eqref{eqn:c2}--\eqref{eqn:calG} coincide with their well-known values for the
ideal gas equation of state when $p_{\infty}=0$, and imply an \emph{a priori} restriction
on any curve fits computed for the model: they must have $\gamma > 1$ in order for the
usual assumptions on $\Gamma$, i.e.\ $\Gamma > 0$, to hold.

Although the four parameters in~\eqref{eqn:stiffened_gas_p}
and~\eqref{eqn:stiffened_gas_T} are treated as constants in practice, in reality they are
weak functions of both pressure and temperature for an actual fluid --- especially far
from phase transition boundaries.  Although various numerical values for the constants have
been proposed previously in the literature~\cite{Metayer_2004}, it appears that no systematic
curve fits (with well-quantified relative error bounds) have thus far been published for the
stiffened gas equation of state for water.

\begin{figure}[hbt]
  \begin{center}
    \includegraphics[width=.75\textwidth]{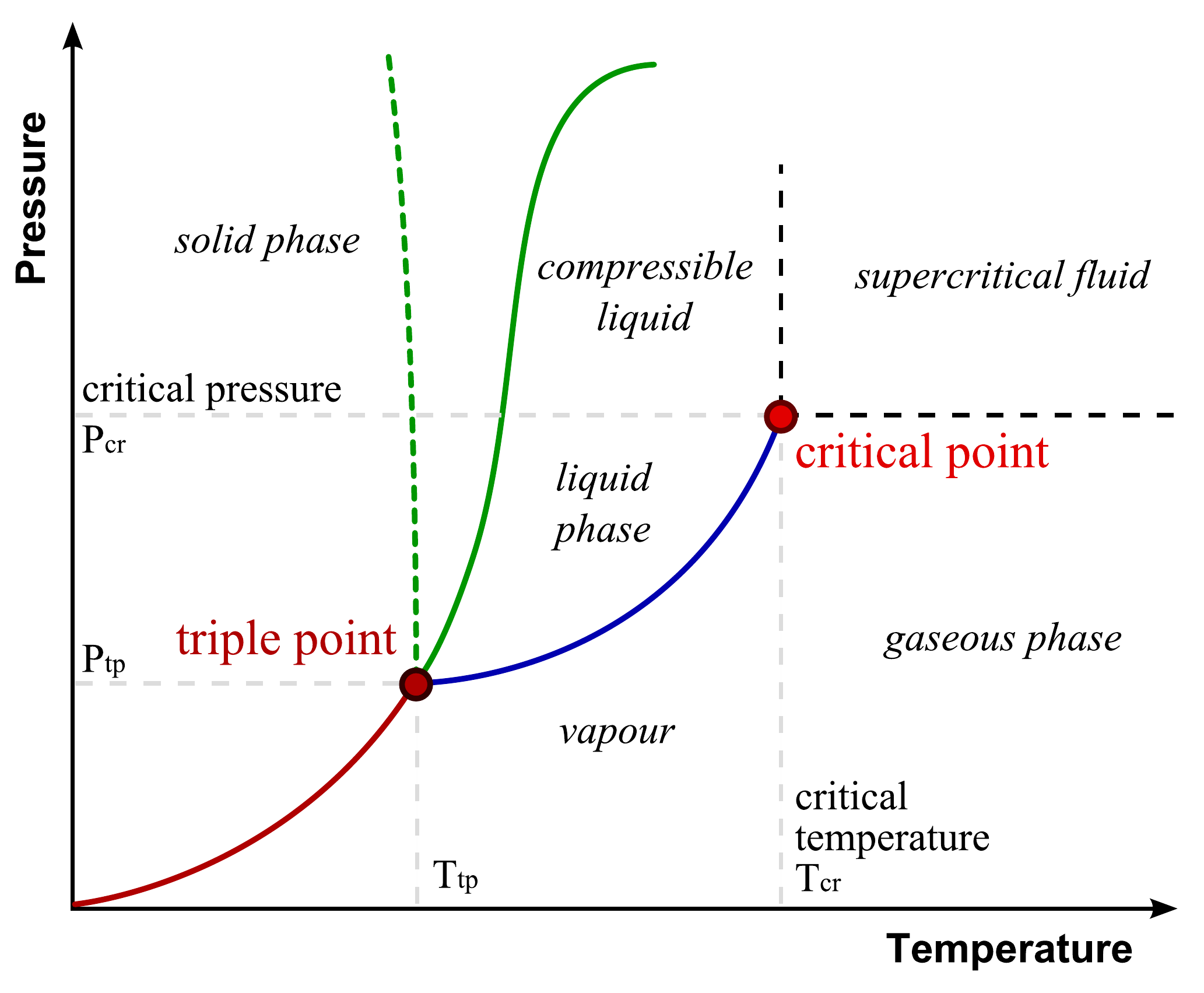}
    \caption{Schematic phase diagram of water~\cite{wiki:phase_diagram}.
      \label{fig:phase_diagram}}
  \end{center}
\end{figure}

A schematic of the basic phase diagram structure of water is given
in Fig.~\ref{fig:phase_diagram}.  The curve fits in this paper are valid only in
the ``compressible liquid'' region of the diagram, i.e.\ for pressures above
the critical pressure (which for water is approximately 22.064 MPa) and below the
critical temperature of approximately 647K.  For lower pressures, especially near the
liquid-vapor transition line, it was found that curve fits based on the stiffened
gas equation of state had unacceptably high relative error levels.  A key observation
is the following: the relative simplicity of the stiffened gas equation of state
leads to a tradeoff in its overall applicability.  If accurate state values near
the phase transition line are required (as is often the case in multi-phase flows)
a more sophisticated model, such as the modified Tait equation of state~\cite{Dymond_1988,Saurel_1999}
should be employed.

The remainder of this paper is organized in the following way: in \S\ref{sec:methodology}, we
describe the functional form of the curve fits, and the normalized linear least-squares
fitting procedure used in computing them.  Then, in \S\ref{sec:results}, the resulting fits
are provided in the form of line plots and tabulated values, along with some discussion of
how they can be utilized effectively.  Finally, in \S\ref{sec:conclusion}, we discuss the
limitations of the stiffened gas equation of state near phase transition boundaries, and
give some suggestions for future work.

\section{Methodology\label{sec:methodology}}
In this section, we describe some general strategies for fitting multiple models
simultaneously, the specific least-squares optimization problems solved in the present work,
and the data gathering and relative error measures used in creating the fits.

\subsection{Fitting Strategy\label{sec:opt_strat}}
Selecting a curve fitting strategy for~\eqref{eqn:stiffened_gas_p}
and~\eqref{eqn:stiffened_gas_T} does not appear to be a straightforward process:
because there are two models that involve a partially-overlapping set
of fit parameters, it is reasonable to ask whether one can find a
\emph{single} set of fit parameters which is \emph{simultaneously}
optimal (in some sense) for both models.  This problem may be
characterized in the framework of ``multi-objective''
optimization~\cite{Steuer_1986}.

Multi-objective optimization algorithms are useful in situations where
improvements in one objective function lead to degradation in one or
more of the other objective functions.  The solution to a given multi-objective optimization
problem is the set of so-called ``Pareto optimal'' or ``Pareto efficient''~\cite{Fudenberg_1983}
points in optimization space: points for which no one objective
function can be improved without making one or more of the other
objective functions worse.

These algorithms typically require the definition of an aggregate
objective function, which is essentially a subjective choice on the
part of the user.  Furthermore, only a single set of parameters can be
selected from among the (possibly large) set of all Pareto optimal
parameter sets, and this choice is, again, subjective.  Multi-objective optimization problems also
frequently employ evolutionary (genetic) algorithms which are
considerably more sophisticated than linear least-squares optimization.

In the present work, we instead employ the following ``decoupled'' fitting
strategy:
\begin{enumerate}

\item {Compute the set $\{\gamma, q, p_{\infty}\}$ for which the
    pressure fit~\eqref{eqn:stiffened_gas_p} is least-squares optimal.}

\item {Using the parameters computed in part 1, compute $c_v$ for which
    the temperature fit~\eqref{eqn:stiffened_gas_T} is least-squares
    optimal.}

\end{enumerate}
The philosophy behind this approach is fairly simple: solution of the
Euler equations (a frequent use-case) requires only a pressure model.
The temperature model is required whenever a thermal conduction term
is present in the energy conservation equation; for example when solving the
full Navier-Stokes equations, or when a wall-heating term is used
to model heat transfer along the length of a pipe in 1D.  Therefore, we make the arbitrary choice
to let the pressure fit be least-squares optimal, and the temperature fit be ``as
good as it can be'' given an optimal pressure fit.  Of course, there
is no reason to expect the temperature curve fits obtained in such a
manner to have an acceptable amount of relative error over the entire fit
range, nevertheless, in \S\ref{sec:results} we will demonstrate
that this strategy is indeed capable of producing reasonable results.

\subsection{Normalized Least-Squares Curve Fits\label{sec:least_squares}}
We can write~\eqref{eqn:stiffened_gas_p} in terms of the internal energy, $e$,
and the specific volume $v \equiv \frac{1}{\rho}$ as
\begin{align}
  \label{eqn:e_model}
  e &= A p v + B v + C
\end{align}
where
\begin{align}
  A &= \frac{1}{\gamma-1}
  \\
  B &= \frac{\gamma p_{\infty}}{\gamma-1}
  \\
  C &= q
\end{align}
Clearly, \eqref{eqn:e_model} demonstrates that the internal energy is a ``degenerate bilinear''
function of $p$ and $v$ (with no term linear in $p$), and is linear in the fit parameters $A$, $B$, and $C$.  Thus it is
theoretically possible to compute $A$, $B$, and $C$ values which produce a least-squares optimal
fit of the ``data'' $e_i$ at specific points $(p_i,v_i)$, for $i=1,\ldots,N$.  The stiffened gas parameters
can then be obtained from $A$, $B$, and $C$ as
\begin{align}
  \label{eqn:gamma_ABC}
  \gamma &= \frac{1}{A} + 1
  \\
  p_{\infty} &= \frac{(\gamma-1)B}{\gamma}
  \\
  \label{eqn:q_ABC}
  q &= C
\end{align}
Unfortunately, attempting to compute fits of~\eqref{eqn:e_model} using non-normalized data
leads to extremely ill-conditioned least-squares matrices, and therefore
unacceptable amounts of round-off error.
We therefore instead consider the least-squares problem of finding $\scaled{A}$, $\scaled{B}$,
and $\scaled{C}$ such that
\begin{align}
  \label{eqn:e_model_normalized}
  \scaled{e} &= \scaled{A} \scaled{p} \scaled{v} + \scaled{B} \scaled{v} + \scaled{C}
\end{align}
is least-squares optimal, where
\begin{align}
  \scaled{e} &\equiv \frac{e - \mean{e}}{\std{e}}
  \\
  \scaled{p} &\equiv \frac{p - \mean{p}}{\std{p}}
  \\
  \scaled{v} &\equiv \frac{v}{\mean{v}}
\end{align}
where $\mean{x}$ and $\std{x}$ are the mean and standard deviation of the variable $x$, respectively.
Note that we arbitrarily select a linear (rather than affine) scaling for the specific volume variable,
since upon expanding~\eqref{eqn:e_model_normalized} we obtain
\begin{align}
  \label{eqn:e_model_normalized_expanded}
  e &= \underbrace{ \frac{\std{e} \scaled{A}}{\std{p} \mean{v}} }_A p v
  + \underbrace{\frac{\std{e}}{\mean{v}}\left( \scaled{B} - \frac{\scaled{A} \mean{p}}{\std{p}} \right)}_B v
  + \underbrace{\left(\mean{e} + \std{e}\scaled{C}\right)}_C
\end{align}
i.e.\ a form identical to the original model~\eqref{eqn:e_model}, with shifted and scaled coefficients.
That is, there is a direct correspondence between the scaled fit
parameters $\scaled{A}$, $\scaled{B}$, and $\scaled{C}$, and the original
fit parameters $A$, $B$, and $C$, which is shown
in~\eqref{eqn:e_model_normalized_expanded}.  Using an affine
transformation for $\scaled{v}$ does not lead to such an isomorphism.  The
actual scalings used are, of course, arbitrary: the same fit will be
produced regardless of the scaling employed.  We merely require that
the scaling improve the conditioning of the least-squares problem
enough to produce sufficiently accurate solutions.

The basic procedure in determining the pressure fit is thus as follows:
\begin{enumerate}
\item Compute $\scaled{A}$, $\scaled{B}$, and $\scaled{C}$ in the normalized variables using~\eqref{eqn:e_model_normalized}
\item Compute $A$, $B$, and $C$ using the relationships in~\eqref{eqn:e_model_normalized_expanded}
\item Compute $\gamma$, $q$, and $p_{\infty}$ using~\eqref{eqn:gamma_ABC}--\eqref{eqn:q_ABC}
\end{enumerate}

Once $\gamma$, $q$, and $p_{\infty}$ are known, the model for $T$ can be written as a function of $e$ and $v$ as
\begin{align}
  \label{eqn:T_fit}
  T = D \left( e - q -p_{\infty} v\right)
\end{align}
where $D \equiv \frac{1}{c_v}$.  A single-parameter least-squares fit can then be used to determine $D$ (and thus $c_v$).
In contrast to the pressure fit, the conditioning of this least-squares problem is not a concern since it reduces to a
single scalar equation.  Therefore we do not pursue a normalized fit for~\eqref{eqn:T_fit} as was done in
the pressure model case, although an isomorphism results under the change of variables $\scaled{T} = \frac{T}{\mean{T}}$.
The lack of ``degrees of freedom'' available for the temperature fit is a direct consequence of the decoupled
approach taken in obtaining the fits.  We observe that the temperature fit will only be accurate if $T \propto e - q -p_{\infty} v$,
a fact that is certainly not guaranteed by the parameters $q$ and $p_{\infty}$ which were optimal for
the pressure fit.

\subsection{Fit Data and Relative Error Measures\label{sec:fit_data}}
The data used in the present work come from the International
Association for the Properties of Water and Steam (IAPWS)
standard~\cite{Saul_1989,Wagner_1995}, which are freely available
online from NIST's
website\footnote{http://webbook.nist.gov/chemistry/fluid}.  Isobaric
data for $25 < p < 300$ MPa (in 5 MPa increments) and $300 < T < 625$
(in 1 Kelvin increments) were obtained from the site and used in
making the fits.  Obviously, the amount and distribution of the data
has a direct effect on the accuracy of the resulting fits.  The
pressure fit parameters were found (via trial and error) not to depend
greatly on the granularity of the pressure data when the increments
were less than about 5 MPa.  Furthermore, away from the phase
transition lines, the internal energy is a nearly linear function of
both pressure and temperature, suggesting that the
models~\eqref{eqn:stiffened_gas_p} and~\eqref{eqn:stiffened_gas_T} are
indeed appropriate.

We use the following relative error measure to judge the suitability of the curve fits produced
by the methods described in this section:
\begin{align}
  \label{eqn:rel_err}
  \varepsilon_p \equiv \frac{\| \bv{p}_m - \bv{p} \|}{\| \bv{p} \|}
\end{align}
where $\bv{p}_m$ is a vector of pressure model values as determined by the fit~\eqref{eqn:e_model},
$\bv{p}$ is the vector of (known) pressure data values at corresponding points, and $\|\cdot \|$ is the discrete $\ell_2$-norm.
An analogous form is used to measure the error in the temperature fits.  We (arbitrarily) consider fits with
less than about 2\% relative error to be suitable for numerical calculations, but of course this
threshold depends on the application in question.  The suitability of
fits is, in general, also subject to other criteria: for example it is
desirable that the fit models not produce negative (non-physical)
values of pressure and temperature for any data points within the
range of applicability.  The fit parameters presented in this paper
all satisfy this criterion.

\section{Results and Discussion\label{sec:results}}

\begin{figure}[!hbt]
  \centering
  \includegraphics[viewport=40 15 800 600,clip=true,width=.85\textwidth]{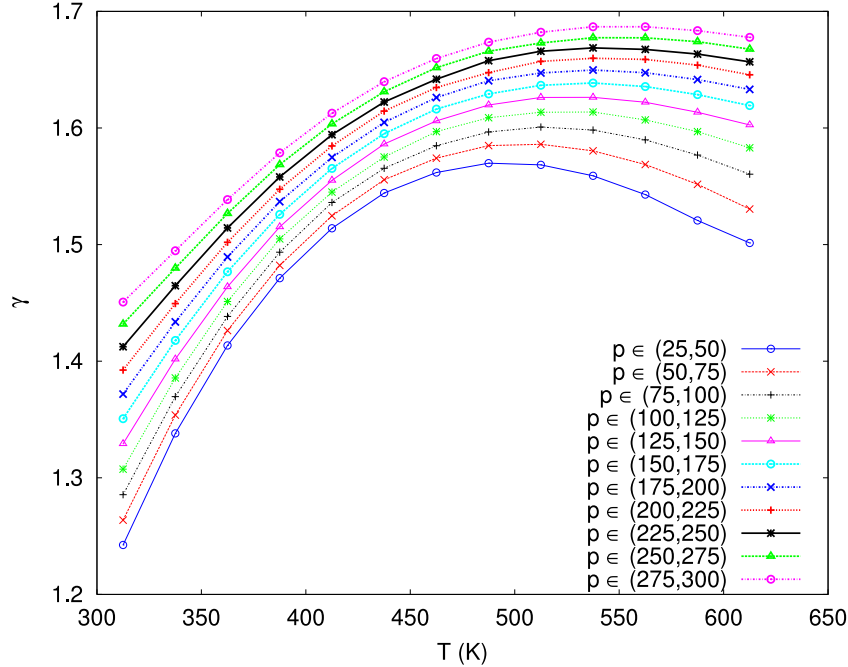}
  \caption{Plots of $\gamma$ (dimensionless) vs.\ temperature for different pressure ranges, all
    pressure ranges are in MPa.\label{fig:parameter_plots_gamma}}
\end{figure}

\begin{figure}[!hbt]
  \centering
  \includegraphics[viewport=40 15 800 600,clip=true,width=.85\textwidth]{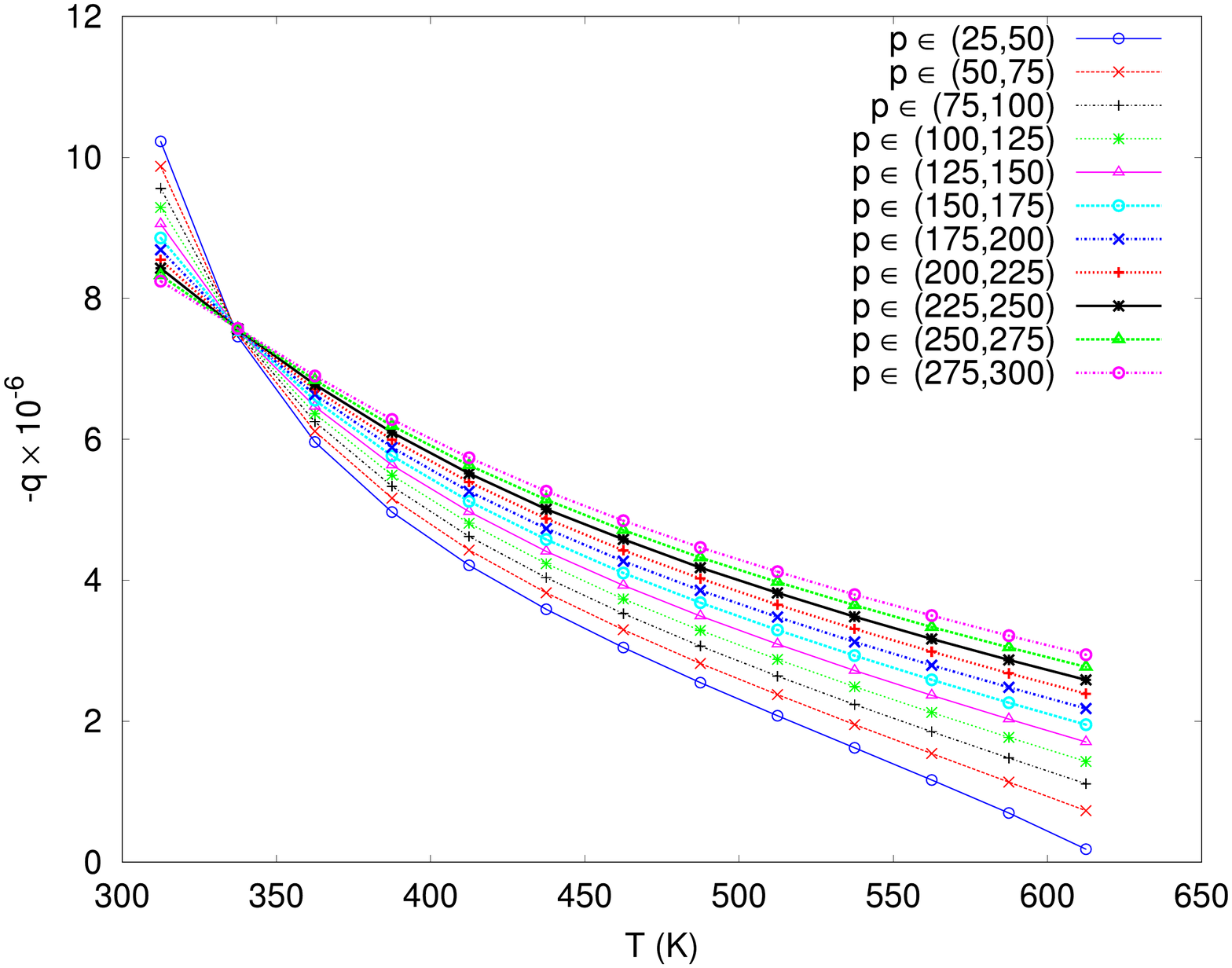}
  \caption{Plots of $-q \times 10^{-6}$ (J/kg) vs.\ temperature for different pressure ranges, all
    pressure ranges are in MPa.\label{fig:parameter_plots_q}}
\end{figure}

\begin{figure}[!hbt]
  \centering
  \includegraphics[viewport=40 15 800 600,clip=true,width=.85\textwidth]{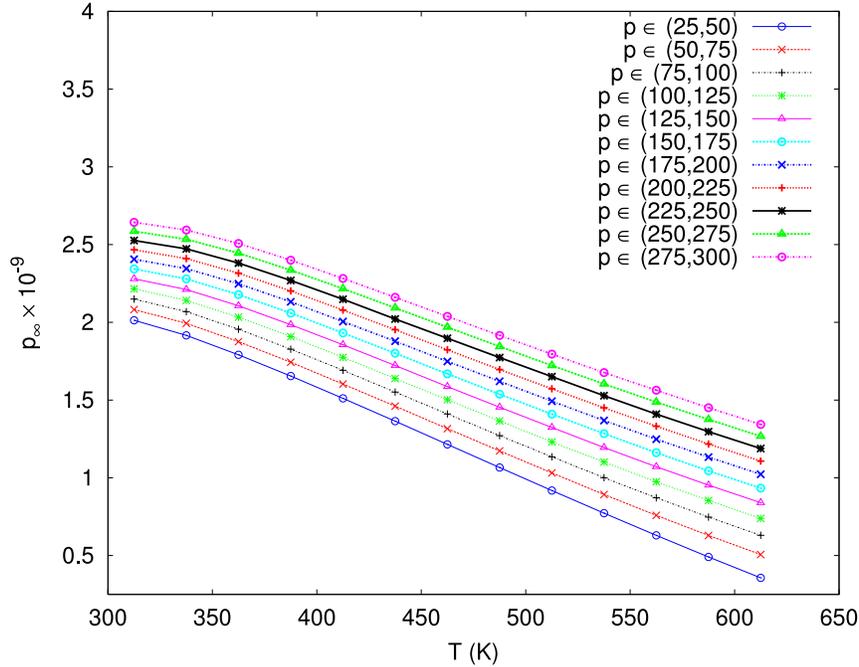}
  \caption{Plots of $p_{\infty} \times 10^{-9}$ (Pa) vs.\ temperature for different pressure ranges, all
    pressure ranges are in MPa.\label{fig:parameter_plots_pinfty}}
\end{figure}

\begin{figure}[!hbt]
  \centering
  \includegraphics[viewport=40 15 800 600,clip=true,width=.85\textwidth]{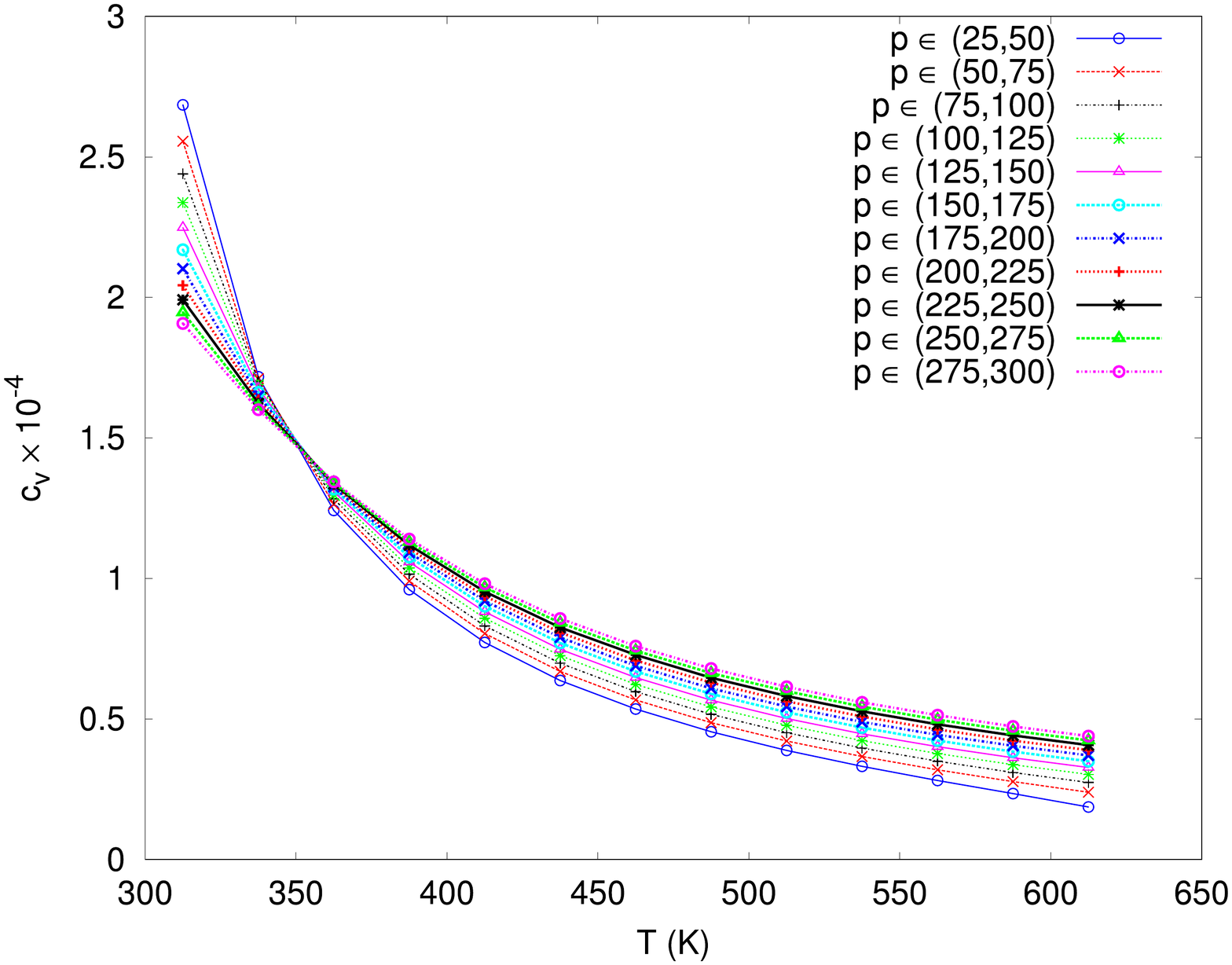}
  \caption{Plots of $c_v \times 10^{-4}$ (J/kg-K) vs.\ temperature for different pressure ranges, all
    pressure ranges are in MPa.\label{fig:parameter_plots_cv}}
\end{figure}

The results of the least-squares curve fitting process described in
\S\ref{sec:methodology} are presented graphically in
Figures~\ref{fig:parameter_plots_gamma}--\ref{fig:parameter_plots_cv}.  The fits themselves were computed
over 25 MPa pressure ranges (from data in 5 MPa increments) and 25
Kelvin temperature ranges (from data in 1 K increments).  Each of the fit parameters is
plotted vs.\ temperature for the various pressure ranges considered.
The figures confirm that each of the optimal fit parameters is indeed a function
of both pressure and temperature over the region of phase space under consideration,
rather than a single constant as posited by the stiffened gas equation of state.


The slopes of the various fit parameter curves have varying levels of dependence
on the pressure level.  The slope of the $p_{\infty}$ curves is almost independent
of the pressure level; for this parameter, increased pressure levels lead to
a constant offset in $p_{\infty}$ over the entire range of
temperatures tested.  For the $q$ and $c_v$ parameters, the pressure effect
is a bit stronger, and increasing
pressure tends to ``flatten'' the curves: $q$ (resp.\ $c_v$) values at
low temperatures decrease with increasing pressure while $q$ (resp.\
$c_v$) values at high temperatures tend to increase with increasing
pressure.  At high temperatures, the $\gamma$ parameter behaves somewhat
differently depending on the pressure. For higher pressures, the $\gamma$
curves ``bend downward'' much less than at lower pressures.  In general, the
fits all improve with increasing pressure, since the fit parameters are more
nearly constant at high pressures.

The total variation in $\gamma$ due to temperature is slightly more
than the amount of variation in $\gamma$ due to pressure.  For
example, in the temperature range $400 < T < 425$, $\gamma$ varies
between approximately 1.54 and 1.64 ($\approx$6\%) over the entire
range of pressure values, but for a given pressure range, say $275 < p
< 300$ MPa, $\gamma$ varies between 1.45 and 1.68 ($\approx$13.7\%). 
The total variations in $c_v$ and $q$ for the lowest pressure range
($25 < p < 50$ MPa, where the highest variation occurs)
over the entire temperature range are considerable:
$93\%$ and $98\%$, respectively.
%
The $q$ and $c_v$ parameters both vary most rapidly at low temperatures, and therefore we
expect the curve fit errors to be overall higher at lower temperatures.  Numerical values for the data plotted in
Figures~\ref{fig:parameter_plots_gamma}--\ref{fig:parameter_plots_cv} are given in Tables~\ref{tab:gamma_table}--\ref{tab:cv_table}.


\begin{figure}[!hbt]
  \centering
    \subfloat[$\varepsilon_T$\label{fig:Terr_of_p}]{
      \includegraphics[viewport=40 15 800 600,clip=true,width=.85\textwidth]{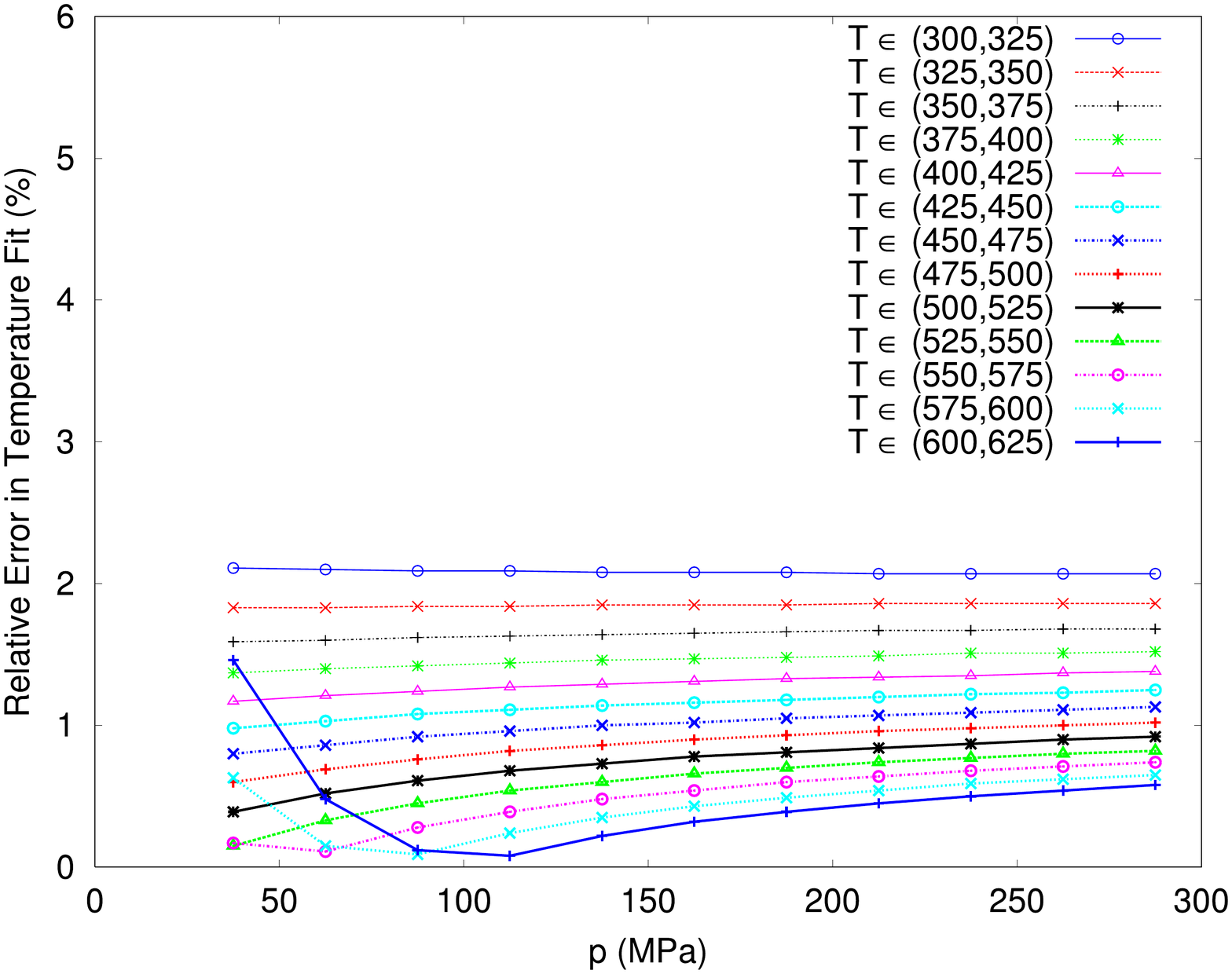}
    }
    \caption{Plots of the relative error $\varepsilon_T$ vs.\ pressure for different temperature ranges, all
      temperature ranges are in K.\label{fig:relerrT_plots}}
\end{figure}

\begin{figure}[hbt]
  \centering
      \includegraphics[viewport=40 15 800 600,clip=true,width=.85\textwidth]{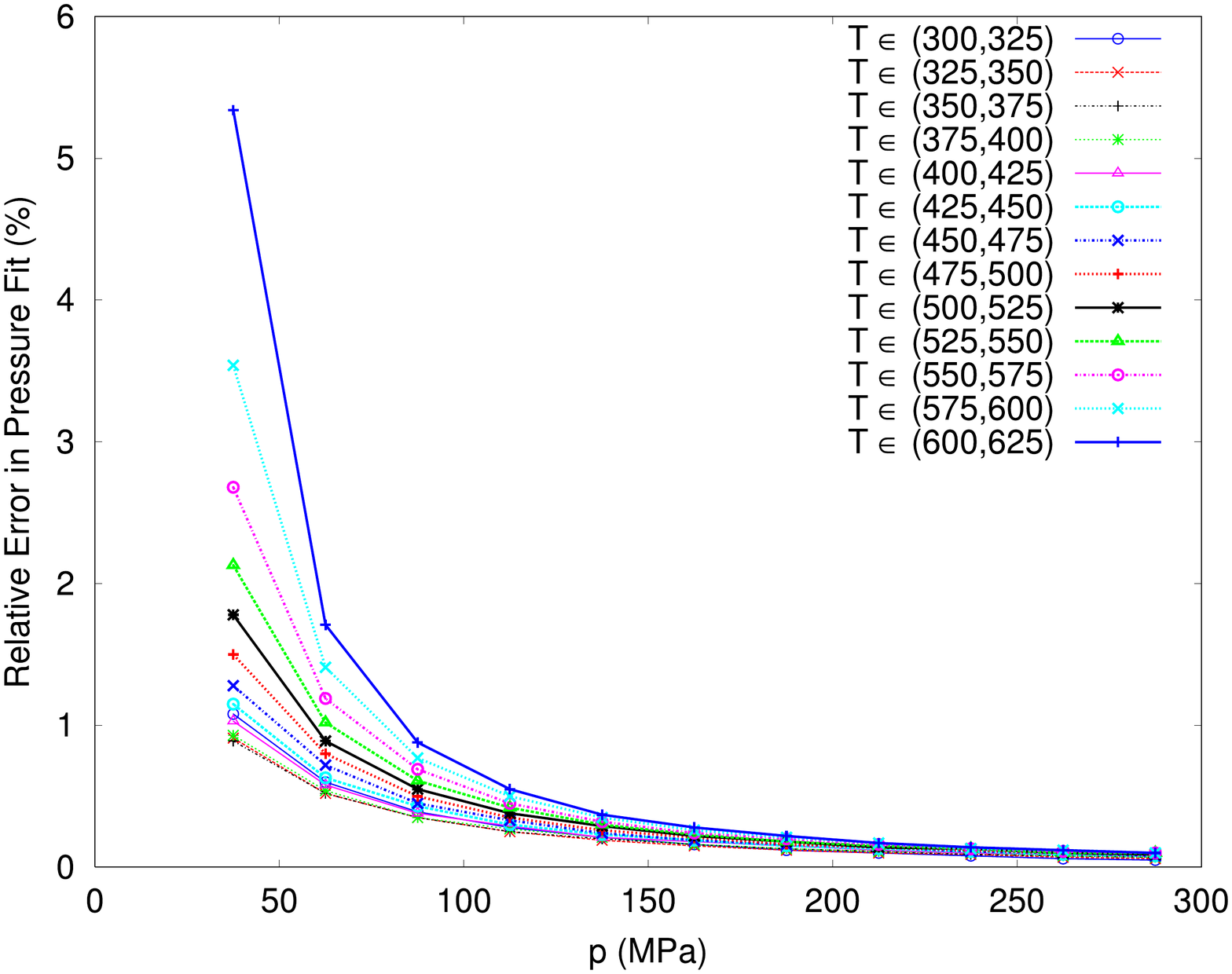}
    \caption{Plots of the relative error $\varepsilon_p$ vs.\ pressure for different temperature ranges, all
      temperature ranges are in K.\label{fig:relerrp_plots}}
\end{figure}

Figures~\ref{fig:relerrT_plots} and~\ref{fig:relerrp_plots} give plots of relative temperature and
pressure errors (as defined by~\eqref{eqn:rel_err}) as functions of
pressure for the different temperature ranges considered.  While the
plots in Figures~\ref{fig:parameter_plots_gamma}--\ref{fig:parameter_plots_cv} were made with respect to
temperature, we found the error plots were easier to understand (and
explain) when plotted with respect to pressure.  The errors in the
temperature fits are less than about 2\% over the entire range of
pressures and temperatures investigated.  The temperature error bounds
improve with increasing temperature, except for the highest
temperatures at the lowest pressures, where the error begins
increasing rapidly.  The latter situation (high temperatures and lower
pressures) corresponds to a region of the phase space diagram which is
closer to the liquid-vapor transition line, and therefore to an
area in which the stiffened gas equation of state is not particularly
well-suited.

The trend of higher error being present in the fits at
lower pressures is also evident in Fig.~\ref{fig:relerrp_plots}, although
in this case the maximum relative error is slightly above 5\%.
The error decreases rapidly at higher pressures, however, which is
in contrast to the nearly-constant error levels seen in the temperature
fits.  We conjecture that this behavior is primarily due to the ``decoupled''
strategy employed in computing the fits (described in
\S\ref{sec:methodology}).  Although our fits only extend to 300 MPa,
we fully expect the trend of low relative errors in the pressure
fits to continue out to much higher pressures.

\section{Conclusion\label{sec:conclusion}}
A method for computing optimal linear least-squares curve fits for the
stiffened gas equation of state over a wide range of pressures and
temperatures was presented, along with relative error bounds for both.  For
high pressures and temperatures, reasonable accuracy in both the
pressure and temperature models is obtained; the curve fits are
unfortunately much less accurate near the liquid-vapor transition
line.  The tabulated parameter values presented here are practical
for use in codes attempting to simulate water as a compressible liquid.

We remark that the data in this paper should \emph{not} be used to
conclude that a more complex equation of state, for example with
pressure- and temperature-dependent parameters, should be utilized in
place of the much simpler stiffened gas equation of state.  Rather, we
recommend that this data be used in the following way: given \emph{a
priori} knowledge of a particular problem, for example operational
pressure and temperature ranges, appropriate constant values of
$\gamma$, $q$, $p_{\infty}$, and $c_v$ should be selected from
Tables~\ref{tab:gamma_table}--\ref{tab:cv_table} and used in the
computation.  Of course, if the operational pressure or temperature
for a given problem exceeds the 25 MPa or 25 K ranges presented in the
tables, some sort of averaging process (based on the tabulated data)
should be employed to select the constant values.

The importance of having accurate equation of state values in fluid
dynamics simulations is difficult to quantify, and of course
problem-dependent.  Parameter sensitivity analysis of nonlinear
partial differential equations is itself an important area of ongoing
research~\cite{Anderson_2005}, and is presently the best
way of determining, for a given problem, how errors in parameter
values translate to errors in problem quantities of interest.
The parametric sensitivity analysis of pressure equation
of state values for the Euler equations is also complicated by
the fact that e.g.\ the flux Jacobians also involve derivatives of
the pressure with respect to the conserved variables.  While such
an analysis is clearly beyond the scope of the present paper, we feel
it represents an interesting avenue of future work.

\section*{Acknowledgments}
The submitted manuscript has been authored by a contractor of the
U.S. Government under Contract DE-AC07-05ID14517. Accordingly, the
U.S. Government retains a non-exclusive, royalty-free license to
publish or reproduce the published form of this contribution, or allow
others to do so, for U.S. Government purposes.


\bibliographystyle{ieeetr}
\bibliography{sgeos_fits}

\begin{thebibliography}{10}

\bibitem{Berry_2010}
R.~A. Berry, R.~Saurel, and O.~L. M\'{e}tayer, ``{The discrete equation method
  (DEM) for fully compressible, two-phase flows in ducts of spatially varying
  cross-section},'' {\em Nuclear Engineering and Design}, vol.~240,
  pp.~3797--3818, Nov. 2010.
\newblock \url{http://dx.doi.org/10.1016/j.nucengdes.2010.08.003}.

\bibitem{Menikoff_1989}
R.~Menikoff and B.~J. Plohr, ``{The Riemann problem for fluid flow of real
  materials},'' {\em Reviews of Modern Physics}, vol.~61, no.~1, pp.~75--130,
  1989.
\newblock \url{http://dx.doi.org/10.1103/RevModPhys.61.75}.

\bibitem{Metayer_2004}
O.~L. M\'{e}tayer, J.~Massoni, and R.~Saurel, ``{Elaborating equations of state
  of a liquid and its vapor for two-phase flow models},'' {\em Int. J. Thermal
  Sciencies (in French)}, vol.~43, pp.~265--276, Mar. 2004.
\newblock \url{http://dx.doi.org/10.1016/j.ijthermalsci.2003.09.002}.

\bibitem{Dymond_1988}
J.~H. Dymond and R.~Malhotra, ``{The Tait equation: 100 years on},'' {\em
  International Journal of Thermophysics}, vol.~9, no.~6, pp.~941--951, 1988.
\newblock \url{http://dx.doi.org/10.1007/BF01133262}.

\bibitem{Saurel_1999}
R.~Saurel, J.~P. Cocchi, and P.~B. Butler, ``{Numerical study of cavitation in
  the wake of a hypervelocity underwater projectile},'' {\em J. of Propulsion
  and Power}, vol.~15, pp.~512--522, July 1999.

\bibitem{wiki:phase_diagram}
Wikipedia, ``Phase diagram,'' 2012.
\newblock \url{http://en.wikipedia.org/wiki/Phase_diagram}, [Online; accessed
  01-Nov-2012].

\bibitem{Steuer_1986}
R.~E. Steuer, {\em {Multiple criteria optimization: theory, computations, and
  application}}.
\newblock New York: John Wiley \& Sons, Inc., 1986.

\bibitem{Fudenberg_1983}
D.~Fudenberg and J.~Tirole, {\em {Game Theory. Chapter 1, Section 2.4}}.
\newblock Cambridge, Mass: MIT Press, 1983.

\bibitem{Saul_1989}
A.~Saul and W.~Wagner, ``{A fundamental equation for water covering the range
  from the melting line to 1273 K at pressures up to 25000 MPa},'' {\em Journal
  of Physical and Chemical Reference Data}, vol.~18, no.~4, pp.~1537--1564,
  1989.
\newblock \url{http://www.nist.gov/data/PDFfiles/jpcrd370.pdf}.

\bibitem{Wagner_1995}
W.~Wagner and A.~Pruss, ``{The IAPWS formulation 1995 for the thermodynamic
  properties of ordinary water substance for general and scientific use},''
  {\em Journal of Physical and Chemical Reference Data}, vol.~31, no.~2,
  pp.~387--535, 2002.
\newblock \url{http://www.teos-10.org/pubs/Wagner_and_Pruss_2002.pdf}.

\bibitem{Anderson_2005}
M.~Anderson, W.~Bangerth, and G.~F. Carey, ``Analysis of parameter sensitivity
  and experimental design for a class of nonlinear partial differential
  equations,'' {\em Int. J. Numer. Meth. Fluids}, vol.~48, pp.~583--605, 2005.
\newblock \url{http://dx.doi.org/10.1002/fld.938}.

\end{thebibliography}

\begin{landscape}
  \begin{table}[hbt]\scriptsize 
  \renewcommand{\tabcolsep}{3pt}
  \centering 
  \caption{Values of $\gamma$ for various pressure (rows) and temperature (columns) ranges.\label{tab:gamma_table}}
    \begin{tabular}{llcccccccccccccc} \toprule
      & \multicolumn{13}{c}{\textbf{Temperature Ranges (K)}} \\
      & & \textbf{300-325} & \textbf{325-350} & \textbf{350-375} & \textbf{375-400} & \textbf{400-425} & \textbf{425-450} & \textbf{450-475}
        & \textbf{475-500} & \textbf{500-525} & \textbf{525-550} & \textbf{550-575} & \textbf{575-600} & \textbf{600-625} \\ \midrule
      \multirow{11}*{\begin{sideways}\textbf{Pressure Ranges (MPa)}\end{sideways}}
        & \textbf{25-50}    & 1.2424 & 1.3381 & 1.4135 & 1.4712 & 1.5140 & 1.5442 & 1.5618 & 1.5698 & 1.5684 & 1.5590 & 1.5429 & 1.5207 & 1.5014 \\
        & \textbf{50-75}    & 1.2637 & 1.3538 & 1.4263 & 1.4822 & 1.5247 & 1.5556 & 1.5741 & 1.5849 & 1.5860 & 1.5804 & 1.5687 & 1.5517 & 1.5305 \\
        & \textbf{75-100}   & 1.2855 & 1.3696 & 1.4384 & 1.4935 & 1.5361 & 1.5654 & 1.5848 & 1.5966 & 1.6008 & 1.5982 & 1.5898 & 1.5768 & 1.5604 \\
        & \textbf{100-125}  & 1.3074 & 1.3856 & 1.4512 & 1.5049 & 1.5450 & 1.5751 & 1.5969 & 1.6088 & 1.6135 & 1.6137 & 1.6069 & 1.5969 & 1.5830 \\
        & \textbf{125-150}  & 1.3292 & 1.4018 & 1.4638 & 1.5152 & 1.5551 & 1.5862 & 1.6061 & 1.6197 & 1.6262 & 1.6263 & 1.6221 & 1.6135 & 1.6026 \\
        & \textbf{150-175}  & 1.3507 & 1.4178 & 1.4767 & 1.5259 & 1.5653 & 1.5951 & 1.6162 & 1.6292 & 1.6366 & 1.6386 & 1.6355 & 1.6286 & 1.6192 \\
        & \textbf{175-200}  & 1.3718 & 1.4337 & 1.4894 & 1.5368 & 1.5748 & 1.6048 & 1.6261 & 1.6405 & 1.6472 & 1.6497 & 1.6474 & 1.6415 & 1.6331 \\
        & \textbf{200-225}  & 1.3924 & 1.4494 & 1.5020 & 1.5475 & 1.5846 & 1.6145 & 1.6347 & 1.6475 & 1.6571 & 1.6598 & 1.6588 & 1.6539 & 1.6456 \\
        & \textbf{225-250}  & 1.4124 & 1.4647 & 1.5143 & 1.5580 & 1.5942 & 1.6223 & 1.6418 & 1.6577 & 1.6658 & 1.6687 & 1.6674 & 1.6634 & 1.6567 \\
        & \textbf{250-275}  & 1.4318 & 1.4798 & 1.5267 & 1.5686 & 1.6035 & 1.6311 & 1.6517 & 1.6658 & 1.6729 & 1.6774 & 1.6773 & 1.6740 & 1.6676 \\
        & \textbf{275-300}  & 1.4507 & 1.4947 & 1.5386 & 1.5787 & 1.6127 & 1.6397 & 1.6596 & 1.6736 & 1.6821 & 1.6868 & 1.6868 & 1.6835 & 1.6777 \\\bottomrule
    \end{tabular}
  \end{table}

  \begin{table}[!h]\scriptsize
    \renewcommand{\tabcolsep}{3pt}
    \centering
    \caption{Table of $-q \times 10^{-6}$ values for various pressure (rows) and temperature (columns) ranges.\label{tab:q_table}}
    \begin{tabular}{llcccccccccccccc} \toprule
      & \multicolumn{13}{c}{\textbf{Temperature Ranges (K)}} \\
      & & \textbf{300-325} & \textbf{325-350} & \textbf{350-375} & \textbf{375-400} & \textbf{400-425} & \textbf{425-450} & \textbf{450-475}
        & \textbf{475-500} & \textbf{500-525} & \textbf{525-550} & \textbf{550-575} & \textbf{575-600} & \textbf{600-625} \\ \midrule
      \multirow{11}*{\begin{sideways}\textbf{Pressure Ranges (MPa)}\end{sideways}}
        & \textbf{25-50}    & 10.229 & 7.4583 & 5.9632 & 4.9684 & 4.2132 & 3.5888 & 3.0462 & 2.5478 & 2.0786 & 1.6228 & 1.1668 & 0.6981 & 0.1848 \\
        & \textbf{50-75}    & 9.8755 & 7.4981 & 6.1147 & 5.1651 & 4.4294 & 3.8227 & 3.2988 & 2.8197 & 2.3782 & 1.9534 & 1.5427 & 1.1368 & 0.7313 \\
        & \textbf{75-100}   & 9.5622 & 7.5243 & 6.2503 & 5.3350 & 4.6227 & 4.0381 & 3.5282 & 3.0660 & 2.6412 & 2.2377 & 1.8526 & 1.4788 & 1.1140 \\
        & \textbf{100-125}  & 9.2909 & 7.5404 & 6.3650 & 5.4912 & 4.8095 & 4.2369 & 3.7354 & 3.2894 & 2.8774 & 2.4897 & 2.1245 & 1.7713 & 1.4302 \\
        & \textbf{125-150}  & 9.0570 & 7.5490 & 6.4663 & 5.6351 & 4.9747 & 4.4108 & 3.9286 & 3.4929 & 3.0951 & 2.7200 & 2.3684 & 2.0318 & 1.7061 \\
        & \textbf{150-175}  & 8.8574 & 7.5549 & 6.5561 & 5.7668 & 5.1212 & 4.5792 & 4.1039 & 3.6825 & 3.2950 & 2.9321 & 2.5882 & 2.2640 & 1.9522 \\
        & \textbf{175-200}  & 8.6898 & 7.5581 & 6.6373 & 5.8862 & 5.2607 & 4.7347 & 4.2708 & 3.8573 & 3.4790 & 3.1262 & 2.7948 & 2.4823 & 2.1806 \\
        & \textbf{200-225}  & 8.5469 & 7.5620 & 6.7110 & 5.9958 & 5.3943 & 4.8756 & 4.4251 & 4.0254 & 3.6531 & 3.3108 & 2.9879 & 2.6815 & 2.3917 \\
        & \textbf{225-250}  & 8.4281 & 7.5648 & 6.7784 & 6.0979 & 5.5164 & 5.0104 & 4.5814 & 4.1800 & 3.8214 & 3.4837 & 3.1681 & 2.8705 & 2.5872 \\
        & \textbf{250-275}  & 8.3278 & 7.5687 & 6.8403 & 6.1928 & 5.6319 & 5.1434 & 4.7126 & 4.3229 & 3.9754 & 3.6472 & 3.3373 & 3.0477 & 2.7709 \\
        & \textbf{275-300}  & 8.2448 & 7.5748 & 6.8996 & 6.2827 & 5.7397 & 5.2651 & 4.8461 & 4.4662 & 4.1225 & 3.7963 & 3.5001 & 3.2141 & 2.9442 \\\bottomrule
    \end{tabular}
  \end{table}

  \begin{table}[hbt]\scriptsize
    \renewcommand{\tabcolsep}{3pt}
    \centering
    \caption{Table of $p_{\infty} \times 10^{-9}$ values for various pressure (rows) and temperature (columns) ranges.\label{tab:pinfty_table}}
    \begin{tabular}{llcccccccccccccc} \toprule
      & \multicolumn{13}{c}{\textbf{Temperature Ranges (K)}} \\
      & & \textbf{300-325} & \textbf{325-350} & \textbf{350-375} & \textbf{375-400} & \textbf{400-425} & \textbf{425-450} & \textbf{450-475}
        & \textbf{475-500} & \textbf{500-525} & \textbf{525-550} & \textbf{550-575} & \textbf{575-600} & \textbf{600-625} \\ \midrule
      \multirow{11}*{\begin{sideways}\textbf{Pressure Ranges (MPa)}\end{sideways}}
        & \textbf{25-50}    & 2.0132 & 1.9161 & 1.7911 & 1.6543 & 1.5109 & 1.3641 & 1.2150 & 1.0659 & 0.9181 & 0.7725 & 0.6299 & 0.4908 & 0.3565 \\
        & \textbf{50-75}    & 2.0822 & 1.9936 & 1.8757 & 1.7433 & 1.6030 & 1.4608 & 1.3162 & 1.1732 & 1.0316 & 0.8925 & 0.7582 & 0.6290 & 0.5064 \\
        & \textbf{75-100}   & 2.1495 & 2.0685 & 1.9556 & 1.8269 & 1.6915 & 1.5514 & 1.4105 & 1.2713 & 1.1350 & 1.0008 & 0.8713 & 0.7472 & 0.6301 \\
        & \textbf{100-125}  & 2.2155 & 2.1409 & 2.0329 & 1.9082 & 1.7750 & 1.6386 & 1.5023 & 1.3657 & 1.2309 & 1.1011 & 0.9746 & 0.8538 & 0.7389 \\
        & \textbf{125-150}  & 2.2800 & 2.2110 & 2.1066 & 1.9845 & 1.8555 & 1.7221 & 1.5867 & 1.4540 & 1.3237 & 1.1944 & 1.0713 & 0.9524 & 0.8398 \\
        & \textbf{150-175}  & 2.3434 & 2.2791 & 2.1785 & 2.0595 & 1.9314 & 1.8011 & 1.6685 & 1.5375 & 1.4096 & 1.2841 & 1.1614 & 1.0446 & 0.9335 \\
        & \textbf{175-200}  & 2.4056 & 2.3453 & 2.2482 & 2.1321 & 2.0050 & 1.8787 & 1.7486 & 1.6207 & 1.4924 & 1.3685 & 1.2482 & 1.1331 & 1.0226 \\
        & \textbf{200-225}  & 2.4667 & 2.4101 & 2.3158 & 2.2017 & 2.0785 & 1.9526 & 1.8232 & 1.6960 & 1.5726 & 1.4504 & 1.3321 & 1.2173 & 1.1076 \\
        & \textbf{225-250}  & 2.5262 & 2.4725 & 2.3810 & 2.2692 & 2.1485 & 2.0212 & 1.8973 & 1.7734 & 1.6506 & 1.5281 & 1.4096 & 1.2967 & 1.1884 \\
        & \textbf{250-275}  & 2.5851 & 2.5338 & 2.4450 & 2.3355 & 2.2164 & 2.0929 & 1.9686 & 1.8440 & 1.7219 & 1.6037 & 1.4869 & 1.3755 & 1.2671 \\
        & \textbf{275-300}  & 2.6430 & 2.5939 & 2.5071 & 2.3995 & 2.2822 & 2.1608 & 2.0383 & 1.9155 & 1.7957 & 1.6767 & 1.5631 & 1.4506 & 1.3432 \\\bottomrule
    \end{tabular}
  \end{table}

  \begin{table}[!h]\scriptsize
    \renewcommand{\tabcolsep}{3pt}
    \centering
    \caption{Table of $c_v \times 10^{-4}$ values for various pressure (rows) and temperature (columns) ranges.\label{tab:cv_table}}
    \begin{tabular}{llcccccccccccccc} \toprule
      & \multicolumn{13}{c}{\textbf{Temperature Ranges (K)}} \\
      & & \textbf{300-325} & \textbf{325-350} & \textbf{350-375} & \textbf{375-400} & \textbf{400-425} & \textbf{425-450} & \textbf{450-475}
        & \textbf{475-500} & \textbf{500-525} & \textbf{525-550} & \textbf{550-575} & \textbf{575-600} & \textbf{600-625} \\ \midrule
      \multirow{11}*{\begin{sideways}\textbf{Pressure Ranges (MPa)}\end{sideways}}
        & \textbf{25-50}    & 2.6854 & 1.7178 & 1.2425 & 0.9604 & 0.7724 & 0.6372 & 0.5356 & 0.4546 & 0.3882 & 0.3316 & 0.2814 & 0.2349 & 0.1871 \\
        & \textbf{50-75}    & 2.5560 & 1.7112 & 1.2644 & 0.9907 & 0.8041 & 0.6692 & 0.5681 & 0.4870 & 0.4219 & 0.3667 & 0.3193 & 0.2774 & 0.2395 \\
        & \textbf{75-100}   & 2.4401 & 1.7015 & 1.2834 & 1.0157 & 0.8314 & 0.6988 & 0.5975 & 0.5168 & 0.4514 & 0.3967 & 0.3502 & 0.3098 & 0.2742 \\
        & \textbf{100-125}  & 2.3380 & 1.6896 & 1.2976 & 1.0381 & 0.8587 & 0.7258 & 0.6231 & 0.5431 & 0.4779 & 0.4231 & 0.3774 & 0.3375 & 0.3028 \\
        & \textbf{125-150}  & 2.2484 & 1.6762 & 1.3092 & 1.0587 & 0.8817 & 0.7482 & 0.6477 & 0.5671 & 0.5018 & 0.4474 & 0.4016 & 0.3623 & 0.3276 \\
        & \textbf{150-175}  & 2.1701 & 1.6628 & 1.3182 & 1.0767 & 0.9014 & 0.7706 & 0.6693 & 0.5895 & 0.5242 & 0.4696 & 0.4234 & 0.3841 & 0.3498 \\
        & \textbf{175-200}  & 2.1025 & 1.6491 & 1.3256 & 1.0922 & 0.9202 & 0.7906 & 0.6896 & 0.6093 & 0.5443 & 0.4898 & 0.4439 & 0.4048 & 0.3705 \\
        & \textbf{200-225}  & 2.0431 & 1.6361 & 1.3316 & 1.1061 & 0.9377 & 0.8082 & 0.7086 & 0.6296 & 0.5633 & 0.5091 & 0.4629 & 0.4234 & 0.3896 \\
        & \textbf{225-250}  & 1.9919 & 1.6235 & 1.3365 & 1.1187 & 0.9534 & 0.8259 & 0.7283 & 0.6469 & 0.5819 & 0.5273 & 0.4812 & 0.4416 & 0.4073 \\
        & \textbf{250-275}  & 1.9469 & 1.6116 & 1.3403 & 1.1298 & 0.9681 & 0.8425 & 0.7434 & 0.6633 & 0.5992 & 0.5443 & 0.4978 & 0.4582 & 0.4238 \\
        & \textbf{275-300}  & 1.9077 & 1.6007 & 1.3440 & 1.1403 & 0.9815 & 0.8575 & 0.7595 & 0.6798 & 0.6149 & 0.5594 & 0.5137 & 0.4738 & 0.4394 \\\bottomrule
    \end{tabular}
  \end{table}

\end{landscape}


\end{document}